\documentclass{article}
\usepackage{graphicx}
\usepackage{amsmath}
\usepackage{graphics}

\input{tcilatex}

\begin{document}

\title{Building multiparticle states with teleportation}
\author{Ernesto Galv\~{a}o$^{1}$ and Lucien Hardy$^{2}$ \\
Centre for Quantum Computation, Clarendon Laboratory,\\
University of Oxford.\\
Parks Road, OX1 3PU Oxford, U.K.\\
\\
$^{1}$\texttt{e.galvao1@physics.ox.ac.uk}\\
$^{2}$\texttt{l.hardy1@physics.ox.ac.uk}}
\maketitle

\begin{abstract}
{\small We describe a protocol which can be used to generate any N-partite
pure quantum state using Einstein-Podolsky-Rosen (EPR) pairs. This protocol
employs only local operations and classical communication between the N
parties (N-LOCC). In particular, we rely on quantum data compression and
teleportation to create the desired state.\ The protocol can be used to
obtain upper bounds for the bipartite entanglement of formation of an
arbitrary N-partite pure state, in the asymptotic limit of many copies.\ We
apply it to a few multipartite states of interest, showing that in some
cases it is not optimal. Generalizations of the protocol are developed which
are optimal for some of the examples we consider, but which may still be
inefficient for arbitrary states.}
\end{abstract}

\section{Introduction}

Quantum entanglement represents a resource that can be used for many
applications, such as teleportation [1], superdense coding [2], quantum
cryptographic key distribution [3] and entanglement enhanced communication
complexity [4]. Much progress has been made in the analysis of bipartite
entanglement of pure and mixed states. The characterization of multiparticle
entangled states is, however, much harder to attain, as some key theorems
that apply to the bipartite states cannot be easily extended to the
multipartite case.

In the case of bipartite pure states a single satisfactory measure of
entanglement has been found [5,6]: the partial von Neumann entropy of one of
the parties. This quantity equals both the states's distillable entanglement
(the number of EPR pairs that can be asymptotically obtained from it using
only classically coordinated local operations) and the state's entanglement
of formation (the number of EPR pairs necessary to asymptotically build the
state with classically coordinated local operations). The characterization
of entanglement in multiparticle states is still an elusive goal. There
seems to be qualitatively different kinds of entanglement in N-partite
states, making it hard to even find higher bounds for the number of
different kinds of entangled states necessary to build N-partite states
irreversibly [7].

In this work we describe a protocol for creating any N-partite pure state $%
\left| \psi\right\rangle $ using only local operations and classical
communication between the N parties (N-LOCC), and a certain number of EPR
pairs. As has been discussed before [8], any measure of entanglement must be
non-increasing under N-LOCC. This will enable us to identify the number of
EPR pairs used as an upper bound for the bipartite entanglement of formation 
$E_{F}(\left| \psi\right\rangle )$. It is likely that multipartite states
will be built from EPR pairs in future practical applications of
multiparticle entanglement; our protocol would then provide bounds for the
efficiency of such a procedure. We also hope that our approach may help
clarify the relations between bipartite and multipartite entanglement, still
a problematic theoretical issue.

Our protocol needs to rely on an asymptotically large number of EPR pairs,
creating a correspondingly large number of copies of the desired N-partite
state $\left| \psi\right\rangle $. $E_{F}(\left| \psi\right\rangle )$ will
then be bounded from above by the number of EPR pairs used per copy of $%
\left| \psi\right\rangle $ generated. We note that in general the procedure
will be irreversible.

\section{Protocol P1}

In this section we describe a protocol that builds a large number of copies
of an arbitrary N-partite pure state using N-LOCC and EPR pairs only. Let us
first see how it would work for an arbitrary 3-qubit pure state $\left|
\psi\right\rangle $.

The basic idea is very simple and consists of distributing the state using a
series of teleportations, keeping track of the amount of EPR pairs used in
each step. Let us suppose that we start with a large number $M$ of copies of 
$\left| \psi\right\rangle $ at Charlie's laboratory. Let us label the three
qubits according to their von Neumann entropies, qubit 1 being the one with
least entropy $S_{1}$, and so on in ascending order. The first step is to do
quantum data compression [9] on the sub-ensembles corresponding to qubits 1
and 2, resulting in two groups of particles asymptotically containing $MS_{1}
$ and $MS_{2}$ maximally compressed qubits.

Now Charlie teleports the set corresponding to qubit 1 to Alice and the
other set to Bob. Thanks to the data compression we performed, these two
teleportation steps will require only $MS_{1}$ EPR pairs shared with Alice
and $MS_{2}$ with Bob. To end the process we decompress the sets at Alice
and Bob, asymptotically recovering $M$ qubits at each party with negligible
error. The resulting state consists of $M$ copies of $\left|
\psi\right\rangle $ distributed among the 3 parties. The above discussion
shows us that an arbitrary 3-partite pure state can be built using 3-LOCC
and a bipartite entanglement of $S_{1}+S_{2}$ EPR pairs, if we consider the
manipulation of a large number of copies of the state.

This protocol can be simply generalized for any number of parties $N>3$. To
see how, let us first recall the sub-additivity property of von Neumann
entropies: if A and B are subsystems of a larger system AB, their entropies
must obey the inequality

\begin{equation*}
S_{A}+S_{B}\geq S_{AB}\text{.}
\end{equation*}
In order to create the N-partite distributed state $\left| \psi\right\rangle 
$ using our protocol we will need to perform $(N-1)$ teleportations. Each
teleportation takes the compressed sub-ensemble corresponding to subsystem $%
\rho_{X}$ (which may include one or more of the final parties' subsystems we
want to distribute) from one party to another, using $S_{X}$ EPR pairs in
the process. Due to the sub-additivity property of entropy, the most
economical series of teleportations will, in general, involve teleporting
larger subparts of $\left| \psi\right\rangle $ first, in order to save on
the number of EPR pairs used in total. As an example, in order to distribute
a state among 4 parties A, B, C and D, it may be advantageous to first
teleport the compressed subsystem CD from site A to site C, and then
teleport subsystems B and D to their respective sites.

We are interested in finding the best series of teleportations from a list
of all possible sequences of $(N-1)$ teleportations that distribute the
N-partite state $\left| \psi\right\rangle $ from a single location. This
optimal series of teleportations can be found through an exhaustive search
and constitutes what we shall call \textbf{protocol P1} for creating $\left|
\psi \right\rangle $; the number of EPR pairs used per copy of state $\left|
\psi\right\rangle $ created will be called \textbf{P1(}$\left| \psi
\right\rangle $\textbf{)} for short. An example of the use of protocol P1
for a simple 4-partite system can be seen in Fig. 1.

\begin{figure}[ptb]
%
{\includegraphics{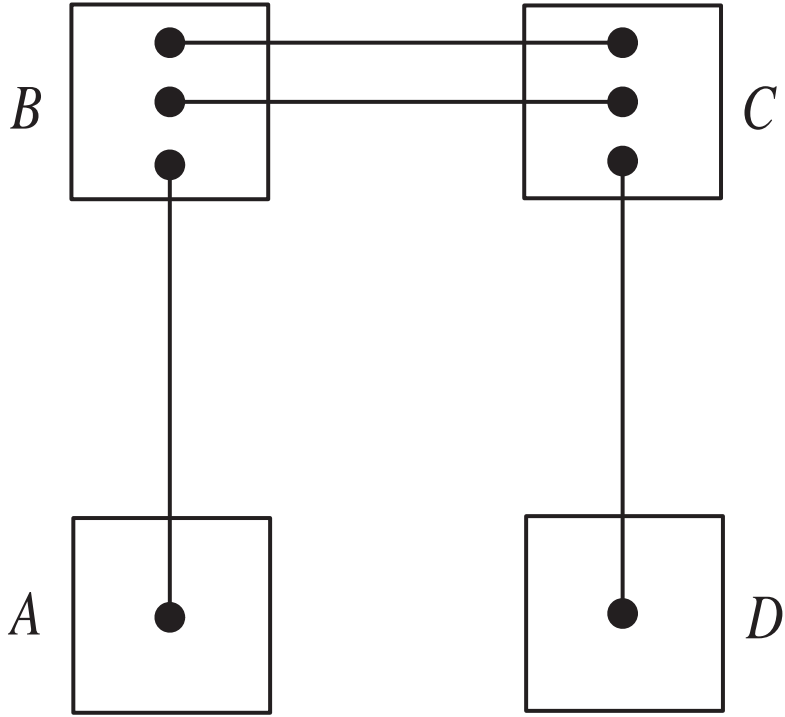}}
\caption{This 4-partite pure state consists of 4 EPR pairs shared between
parties A, B, C and D. A naive approach would involve teleporting each
party's sub-system from party C, using a total of $S_{A}+S_{B}+S_{D}=1+3+1=5$
EPR pairs per copy. Protocol P1 invokes the most economical sequence of
teleportations, one of which in this case is: first subsystem AB from C, and
then the other 2 teleportations, spending the optimal amount of $%
S_{AB}+S_{A}+S_{D}=2+1+1=4$ EPR pairs.}
\label{fig1}
\end{figure}

\section{Results for some classes of states}

In this section we calculate $P1(\left| \psi\right\rangle )$ for some
multipartite states of interest. We will see that protocol P1 is not
optimal, at least for some classes of states. $P1(\left| \psi\right\rangle )$
will then represent an upper bound for $E_{F}(\left| \psi\right\rangle )$. $%
P1(\left| \psi\right\rangle )$ will also be an upper bound for the amount of
bipartite entanglement distillable from $\left| \psi\right\rangle $, as this
quantity is bounded from above by $E_{F}$.

The tripartite case is especially simple: if the three sub-systems of $%
\left| \psi\right\rangle $ have entropies $S_{1},S_{2}$ and $S_{3}$
(labeling the systems such that $S_{1}\leq S_{2}\leq S_{3}$ ), then $%
P1(\left| \psi\right\rangle )=$ $S_{1}+S_{2}$ EPR pairs. In the case of an
N-partite state $\left| \phi\right\rangle $ ($N>3$) there is not such a
closed formula for P1($\left| \phi\right\rangle $); as we discussed above,
in order to find P1($\left| \phi\right\rangle $) we will need to find the
most economical series of $(N-1)$ teleportations that will distribute $%
\left| \phi \right\rangle $.

\subsection{N-GHZ states}

Let us apply protocol P1 to build a generalized N-GHZ state

\begin{equation}
\left| N-GHZ\right\rangle =\frac{1}{\sqrt{2}}\left( \left| 0^{\otimes
N}\right\rangle +\left| 1^{\otimes N}\right\rangle \right) ,
\end{equation}
which is a maximally entangled state of N qubits, also known as a N-cat
state. If X is a non-trivial subset of the parties, then the reduced density
matrix of subset X is defined as

\begin{equation}
\rho_{X}(\left| \psi\right\rangle )=Tr_{\overline{X}}(\left| \psi
\right\rangle \left\langle \psi\right| ).
\end{equation}

The N-GHZ states have the property that for every non-trivial subset X we
have $S_{X}(\left| \psi\right\rangle )=1$. This means that we use exactly 1
EPR pair in each teleportation step, no matter which subsystem of $\left|
N-GHZ\right\rangle $ we choose to teleport. Therefore, all sequences of
teleportations that distribute the state $\left| N-GHZ\right\rangle $ will
use exactly the same amount of EPR pairs, this number being

\begin{equation}
P1(\left| N-GHZ\right\rangle )=(N-1).
\end{equation}
Since all teleportation sequences are equivalent, we may as well choose one
in particular: let us consider the one that distributes each subsystem
directly to its definitive party, thus teleporting $(N-1)$ qubits from a
single location. Since each qubit is in a maximally mixed state, the quantum
data compression would be of no use, making it unnecessary for us to operate
on multiple copies of the state.

This establishes the correspondence between protocol P1 and the one by
Zeilinger \textit{et al. }[10] for building a single copy of a N-GHZ state
of N qubits from $(N-1)$ EPR pairs. Protocol P1 is as efficient as
Zeilinger's, but has a different physical meaning: we are simply teleporting
each party's subsystem from a single location, using a total of $(N-1)$ EPR
pairs.

The above discussion generalizes trivially for the case of N-GHZ states in
which each party possesses k-level subsystems instead of qubits.
Incidentally, the N-GHZ states represent the worst case scenario for
protocol P1, in the sense that a single copy of \textit{any} N-partite state
of N qubits can be prepared locally and distributed using $P1(\left|
N-GHZ\right\rangle )$ EPR pairs. It is not known whether P1 is optimal for
N-GHZ states.

\subsection{\label{Schmidt}N-partite Schmidt decomposable states}

These are states which can be written in Schmidt form

\begin{equation}
\left| \psi^{ABC...N}\right\rangle =\sum_{i}a_{i}\left| i^{A}\right\rangle
\left| i^{B}\right\rangle \left| i^{C}\right\rangle ...\left|
i^{N}\right\rangle ,
\end{equation}
where the $\left| i^{X}\right\rangle $ represent an orthonormal basis in
party X's Hilbert space. Note that the coefficients $a_{i}$ may be made to
be real by convenient redefinition of the basis vectors. For these states
all reduced density matrices $\rho_{X}$ have the same von Neumann entropy of 
$S_{Schmidt}=\sum_{i}a_{i}^{2}\log_{2}a_{i}^{2}$, and therefore

\begin{equation}
P1(\left| \psi^{ABC...N}\right\rangle )=(N-1)S_{Schmidt}.
\end{equation}

Bennett \textit{et al.}\ [7] showed that a state $\left|
\psi^{ABC...N}\right\rangle $ can (in the asymptotic limit of many copies)
be transformed reversibly by N-LOCC into a set of $S_{Schmidt}$ N-GHZ states
of N qubits. Thus we may achieve the creation of $\left|
\psi^{ABC...N}\right\rangle $ by two different routes: either using P1
directly or first creating $S_{Schmidt}$ N-GHZ states with P1 and then using
Bennett \textit{et al.}'s procedure to turn them into $\left|
\psi^{ABC...N}\right\rangle $. We have seen above that $P1(\left|
N-GHZ\right\rangle )=(N-1)$ EPR pairs, which means that the two routes
demand the same number of EPR pairs for creating the state. In view of this,
we see that P1's optimality for creating N-GHZ states (if it can be proved)
would mean P1 would be an optimal protocol for Schmidt decomposable
N-partite states as well.

\subsection{\label{sec sym epr}Symmetrical N-partite states of $\binom{N}{2}$
EPR pairs}

Let us consider the N-partite state $\left| N-Toast\right\rangle $
consisting of $\binom{N}{2}=\frac{N(N-1)}{2}$ shared EPR pairs, one between
each pair of parties. This state has been named a N-Toast state after the
custom of clinking glasses during a toast [7]. We choose these states here
as an example of the inefficiency of P1 for some classes of states.

The most economical series of teleportations that builds $\left|
N-Toast\right\rangle $ consists again of teleporting each party's subsystem
directly from an initial party. In order to see this, let us calculate the
entropies of the reduced density matrices corresponding to all non-trivial
sets of parties $X$.

Let $\rho_{X}$ denote the reduced density matrix corresponding to an
arbitrary set of M parties ($1\leq M\leq N-1$). Given this partition of the
set of parties, the $\binom{N}{2}$ EPR pairs in state $\left|
N-Toast\right\rangle $ can be grouped in three categories: $\binom{M}{2}$
shared between parties in $X$; $\binom{N-M}{2}$ shared between parties in $%
\overline{X}$; and $M(N-M)$ shared between a party in $X$ and another in $%
\overline{X}$. The trace operation over the $N-M$ parties will result in a $%
\rho_{X}$ that is a tensor product of the $\binom{M}{2}$ EPR pairs within $X$
and the identity matrix corresponding to the EPR pairs that were `broken'
between $X$ and $\overline{X}$. Thus we have

\begin{equation}
\rho_{X}=\frac{\mathbf{I}}{2^{M(N-M)}}\otimes\left| \binom{M}{2}\text{ EPR's}%
\right\rangle \left\langle \binom{M}{2}\text{ EPR's}\right| .   \label{rox}
\end{equation}

The entropy of $\rho_{X}$ can then be easily evaluated to be $S_{X}=M(N-M)$. 
$S_{X}$ corresponds to the number of EPR pairs necessary to teleport
subsystem $\rho_{X}$ as part of protocol P1. Since P1 consists of $N-1$
teleportations, we cannot hope to perform P1 and use less than

\begin{equation*}
(N-1)\min\{S_{X}\}=(N-1)^{2}
\end{equation*}
EPR pairs (where the minimum is taken over all possible $X$). This is
actually achievable, as we may teleport each party's subsystem directly from
an initial one, using a total number of EPR pairs equal to

\begin{equation}
P1(\left| N-Toast\right\rangle )=(N-1)^{2}.
\end{equation}
The fact that state $\left| N-Toast\right\rangle $ consists of a collection
of $\binom{N}{2}$ EPR pairs makes it obvious that $E_{F}(\left|
N-Toast\right\rangle )=\binom{N}{2}$, which is clearly below $P1(\left|
N-Toast\right\rangle )$. For these states

\begin{equation}
\frac{P1(\left| N-Toast\right\rangle \text{)}}{E_{F}(\left|
N-Toast\right\rangle )}=\frac{(N-1)^{2}}{\binom{N}{2}}=\frac{2(N-1)}{N},
\end{equation}
which shows that our protocol is particularly inefficient for $\left|
N-Toast\right\rangle $ states --\ $P1(\left| N-Toast\right\rangle )$ is as
high as twice the actual bipartite entanglement of formation for large $N$.

\section{Other bounds for $E_{F}$}

We have just seen that protocol P1 is not optimal. This inefficiency prompts
us to obtain bounds for $E_{F}$ using other methods, as a way of assessing
the protocol's shortcomings. This can be done for the specific case of N-GHZ
states by appealing to a recent theorem proved by Julia Kempe [11]:

\textbf{Definition}: \textit{Two N-partite states are said to be N-LOCC
incommensurate when they cannot be transformed into each other either way by
any N-LOCC protocol.}

\textbf{Theorem} (Kempe):\textit{\ Two N-partite states }$\left|
\psi\right\rangle $\textit{\ and }$\left| \phi\right\rangle $\textit{\ are
N-LOCC incommensurate if and only if they are not equivalent under local
unitary transformations at each party.}

Kempe applied this theorem to prove that $(N-1)$ N-GHZ states and an $\left|
N-Toast\right\rangle $ state are N-LOCC incommensurate. In both cases the
density matrices of each party are identical and equal to

\begin{equation}
\rho=\frac{\mathbf{I}}{2^{(N-1)}}.
\end{equation}
The $\left| N-Toast\right\rangle $ state has inseparable $\rho_{X}$ for any
choice of X other than a single party (see eq. (\ref{rox})). The state
consisting of $(N-1)$ N-GHZ's has separable $\rho_{X}$ for all choices of X.
Using Kempe's theorem and the fact that local unitaries cannot increase
entanglement between parties, we prove that the states under consideration
are N-LOCC incommensurate. A similar result was also obtained by Bennett 
\textit{et al.} in [7].

We will now argue that this result holds true for exact transformations
between arbitrarily large numbers of copies of N-GHZ and EPR states. Let us
consider the two following N-partite systems: the state $\left|
GHZ\right\rangle $ consisting of a tensor product of $k(N-1)$ N-GHZ states
of N qubits, each party possessing a subsystem with $k(N-1)$ qubits in state

\begin{equation}
\rho_{GHZ}=\frac{\mathbf{I}}{2^{k(N-1)}};
\end{equation}
and state $\left| Toasts\right\rangle $ consisting of the tensor product $%
\left| N-Toast\right\rangle ^{\otimes n}$, with each party's subsystem in
state

\begin{equation}
\rho_{Toasts}=\frac{\mathbf{I}}{2^{n(N-1)}}.
\end{equation}
We will have equal density matrices at each party for $\left|
GHZ\right\rangle $ and $\left| Toasts\right\rangle $ if $k/n=1,$ making it
possible for us to apply the same reasoning as we did above for the case
with 2 GHZ and 3 EPR states. Thus we have established that $\left|
N-GHZ\right\rangle ^{\otimes k(N-1)}$ and $\left| N-Toast\right\rangle
^{\otimes n}$ are incommensurate states. In particular, this means that we
need strictly more than $\binom{N}{2}/(N-1)=N/2$ EPR pairs to create one
N-GHZ state, no matter how clever our N-LOCC protocol is. In all of the
above we have been considering exact transformations between the states. In
references [5,7] the authors define larger equivalence classes between
N-partite states, allowing for slightly imperfect fidelity $(1-\varepsilon)$
in the transformation and for $k/n$ that deviates from 1 by a small quantity 
$\delta $. The question of whether the bound of $N/2$ EPR pairs per N-GHZ
state created holds for the transformations considered in [5,7] is open.

Combining the lower bound of $N/2$ EPR pairs per N-GHZ with the higher bound
given by protocol P1, we can write:
\begin{equation}
\frac{N}{2}<E_{F}(\left| N-GHZ\right\rangle )\leq N-1. 
\label{bound for ghz}
\end{equation}
It is interesting to note that the above result implies that there is a
fundamental irreversibility when we transform N-partite entanglement into
bipartite and vice-versa, using exact transformations of a large number of
copies. For example, a 3-GHZ cannot be asymptotically converted into $3/2$
of an EPR pair, yet we need strictly more than $3/2$ of an EPR pair to
asymptotically recreate a 3-GHZ.

P1 is obviously optimal for a trivial set of states, such as that on Fig. 1.
It is, however, inefficient for states like that of section \ref{sec sym epr}
above. For N-GHZ states, as far as we know, there is no known protocol more
efficient than P1, and inequality (\ref{bound for ghz}) represents what
seems to be the best bounds as yet for their bipartite entanglement of
formation. If it can be proved that P1 is optimal for N-GHZ\ states, then it
would be optimal for Schmidt decomposable N-partite states as well, as we
discussed in section \ref{Schmidt} above.

\section{Other protocols}

Protocol P1 is not the most general protocol using quantum data compression
and teleportation to build multipartite states.\ In this section we will
briefly describe three successive generalizations. These generalizations all
involve taking advantage of the fact that each party's system may be\
considered as consisting of subsystems which can be compressed and
teleported independently.

First we describe protocol P2. Take an N-partite state in the Hilbert space $%
\mathcal{H}_{A}\otimes\mathcal{H}_{B}\otimes\mathcal{H}_{C}\otimes ...\otimes%
\mathcal{H}_{Z}$ (where $A,B,...Z$ are the parties).\ Now, it may be the
case that some or all of the spaces $\mathcal{H}_{A}$,$\mathcal{H}_{B},...%
\mathcal{H}_{Z}$ can themselves be written as tensor products of smaller
spaces. For example, we may write the space $\mathcal{H}_{A}$ as a tensor
product of $m$ spaces

\begin{equation*}
\mathcal{H}_{A}=\mathcal{H}_{A}^{1}\otimes\mathcal{H}_{A}^{2}\otimes
...\otimes\mathcal{H}_{A}^{m}. 
\end{equation*}
Writing the Hilbert space in this way corresponds to dividing each party's
system into subsystems.\ The most general way of doing this for a party $A$
is the following.\ We first write $\dim(\mathcal{H}_{A})$ as the product of
its prime factors. If there are $m$ such prime factors then we can regard $A$
as consisting of $m$ subsystems, each having the dimension of one of the
prime factors. All possible subsystems of system $A$ consist either of these 
$\ m$ prime-dimensional subsystems, or of the composite-dimensional
subsystems obtained by grouping these in different ways.\ To obtain the most
general way of identifying subsystems we can consider performing unitary
transformations at each party, either passively (on the Hilbert space) or
actively on the system itself (in which case we would need to undo the
transformation at a later stage).

Now let us see the steps that constitute protocol P2. We start by preparing
many copies of the state we wish to build at one location. Then we consider
all possible ways in which the systems $A,B,...Z$ (which must be distributed
to each party) can be decomposed in a tensor product of prime-dimensional
subsystems.\ The idea now is to analyze all possible ways in which these
subsystems, or groups of them, can be successively compressed at one party,
teleported and decompressed at another party. Protocol P2 is taken to be the
most efficient way of building the N-partite state by such a series of
teleportations of subsystems.

This protocol will be optimal for any state that consists of a tensor
product of shared bipartite states, in particular for $N$-Toast states.\
This becomes clear if we consider a simple example. Take a 3-Toast state

\begin{align}
\left| 3-Toast\right\rangle & =\frac{1}{\sqrt{2}}(\left|
0_{A1}0_{B1}\right\rangle +\left| 1_{A1}1_{B1}\right\rangle )\otimes \\
& \frac{1}{\sqrt{2}}(\left| 0_{A2}0_{C1}\right\rangle +\left|
1_{A2}1_{C1}\right\rangle )\otimes  \notag \\
& \frac{1}{\sqrt{2}}(\left| 0_{B2}0_{C2}\right\rangle +\left|
1_{B2}1_{C2}\right\rangle ).  \notag
\end{align}
We have seen that P1 is inefficient at creating N-Toast states; P1 would
take 4 EPR pairs to create this tensor product of 3 EPR pairs.\ We write
system $A$ as being composed of two 2-dimensional subsystems $A_{1}$ and $%
A_{2}$, $B$ composed of subsystems $B_{1}$ and $B_{2}$, and $C$ of $C_{1}$%
and $C_{2}$.\ This state can be built in three teleportation steps.\ We
start with many copies of $\left| 3-Toast\right\rangle $ at location $A.$
First we teleport system $C_{1}$ to location $C$ (since subsystem $C_{1}$ is
maximally mixed there is no need to compress it).\ Then we take subsystems $%
B_{1},B_{2}$ and $C_{2}$ and, regarding them as one subsystem, we compress
and teleport them to location $B$, where they can be decompressed.\ Finally,
we teleport subsystem $C_{2}$ from location $B$ to location $C$ (again,
there is no need for compression). Each of these three steps requires one
EPR pair per copy of $\left| 3-Toast\right\rangle $ and hence this procedure
is optimal, constituting an example of P2.\ It is clear that this protocol
will be optimal for any state consisting of shared bipartite states.

Protocol P2 reduces to P1 in the case where all systems $A,B,...Z$ have a
prime number of dimensions (thus being not decomposable).\ This observation
motivates protocol P3.\ In protocol P3 we allow the introduction of
ancillary systems $R_{A},R_{B},...R_{Z}$ prepared in some known state. Let
system $A^{\prime}$ consist of $A$ and $R_{A},$ and similarly for $B^{\prime
},C^{\prime},...Z^{\prime}$.\ Protocol P3 consists of applying P2 to the
primed systems and optimizing also over all possible ancilla choices.\ The
introduction of ancillae allows for greater flexibility of manipulation of
the prime-dimensional subsystems, as unprimed systems with prime dimension
may result in primed decomposable systems.

It is easy to find examples of states for which protocol P3 is more
efficient than both P1 and P2. For example, let us consider the following
tripartite state closely resembling the $\left| 3-/Toast\right\rangle $
state, but in a 5x5x5 dimensional Hilbert space:

\begin{align}
\left| \phi\right\rangle & =\sqrt{1-\varepsilon}\left| 3-Toast\right\rangle +%
\sqrt{\varepsilon}\left| 4,4,4\right\rangle \\
& =\sqrt{1-\varepsilon}\frac{1}{\sqrt{8}}\left[ 
\begin{array}{c}
\left| 0,0,0\right\rangle +\left| 0,1,1\right\rangle +\left|
1,0,2\right\rangle +\left| 1,1,3\right\rangle \\ 
+\left| 2,2,0\right\rangle +\left| 2,3,1\right\rangle +\left|
3,2,2\right\rangle +\left| 3,3,3\right\rangle
\end{array}
\right]  \notag \\
& +\sqrt{\varepsilon}\left| 4,4,4\right\rangle ,  \notag
\end{align}
where we have rewritten the $\left| 3-Toast\right\rangle $ state in a
different form, adding a term proportional to $\sqrt{\varepsilon}$ and
orthogonal to the space spanned by the $\left| 3-Toast\right\rangle $ state.

Eigenvalues of density matrices that are continuous functions of a parameter 
$\varepsilon $ are themselves continuous functions of $\varepsilon $ [12].
Hence, the entropies of the partial density matrices obtained from $\rho
_{\phi }$ will differ from those of $\left| 3-Toast\right\rangle
\left\langle 3-Toast\right| $ by arbitrarily small amounts, for suitable
values of $\varepsilon $. We are now dealing with subsystems in a
prime-dimensional (5) Hilbert space, which means that we are unable to
identify subsystems in each party's system and apply P2. Therefore

\begin{equation*}
P2(\left| \phi\right\rangle )=P1(\left| \phi\right\rangle )=S(\rho
_{A})+S(\rho_{B}) 
\end{equation*}
which is arbitrarily close to 4, for small $\varepsilon$.

Protocol P3 improves on this by adding a 3-qubit ancilla to each 5-level
system. We then map the state of each 5-level system into its ancilla
through the unitary operation that takes (for Alice's system)

\begin{align*}
\left| 0_{Alice}\right\rangle \left| 0_{R1},0_{R2},0_{R3}\right\rangle &
\rightarrow\left| 0_{Alice}\right\rangle \left|
0_{R1},0_{R2},0_{R3}\right\rangle , \\
\left| 1_{Alice}\right\rangle \left| 0_{R1},0_{R2},0_{R3}\right\rangle &
\rightarrow\left| 0_{Alice}\right\rangle \left|
0_{R1},1_{R2},0_{R3}\right\rangle , \\
\left| 2_{Alice}\right\rangle \left| 0_{R1},0_{R2},0_{R3}\right\rangle &
\rightarrow\left| 0_{Alice}\right\rangle \left|
1_{R1},0_{R2},0_{R3}\right\rangle , \\
\left| 3_{Alice}\right\rangle \left| 0_{R1},0_{R2},0_{R3}\right\rangle &
\rightarrow\left| 0_{Alice}\right\rangle \left|
1_{R1},1_{R2},0_{R3}\right\rangle , \\
\left| 4_{Alice}\right\rangle \left| 0_{R1},0_{R2},0_{R3}\right\rangle &
\rightarrow\left| 0_{Alice}\right\rangle \left|
0_{R1},0_{R2},1_{R3}\right\rangle ,
\end{align*}
and similarly for Bob's and Charlie's systems. This amounts basically to
increasing the dimensionality of each system from 5 to 8, as we may discard
the original systems and deal with the ancillae instead. Since 8 is a
composite number, now we can apply protocol P3, which consists of finding
the optimal way of distributing this 9-qubit system among Alice, Bob and
Charlie through teleportation. It suffices to say that the optimal way will
be similar to the one we adopted when we applied protocol P2 to the $\left|
3-Toast\right\rangle $ state, with the difference that now we have one extra
qubit at each party (like ancillary bit $R3$ in the example above) which is
minimally entangled with the rest of the system. By the same continuity
argument we see that P3($\left| \phi\right\rangle $) is as close as we like
to P2($\left| 3-Toast\right\rangle $)$=3$, for small $\varepsilon$. This
provides us with an example for which protocol P3 is better than both P1 and
P2.

The introduction of the ancillae in protocol P3 was not done in the most
general way possible. We considered local operations at each party that
entangled ancillae to the systems corresponding to a single copy of the
state we want to prepare. As we have argued, this allows for a more general
manipulation of the state, and perhaps to a more efficient series of
teleportations that build it. We may, however, operate with a large number
of copies of the state, introducing ancillae which couple to a certain
number $g$ of copies, instead of a single one.\ Then our basic units would
consist of one ancilla for each set of $g$ copies of system $A$, one for
each $g$ copies of system $B$, and so on up to the last system $Z$.\
Protocol P4 can then be defined as the application of P2 to these larger
units, optimizing over $g$ and ancilla sizes. P4 is more general than P3 and
may be more efficient for some states.

\section{Conclusions}

We have described a protocol that can be used (in the asymptotic limit of
many copies) to build any N-partite pure quantum state. Protocol P1 uses
quantum data compression and teleportation to distribute a large number of
copies of the state among the N parties, consuming bipartite entanglement in
the form of EPR pairs.

The efficiency of the protocol depends on the state we want to create. It is
optimal for some simple states (such as that in Fig. 1), but can be shown to
be inefficient for N-Toast states, discussed in sec. \ref{sec sym epr}. Its
efficiency for arbitrary states is hard to evaluate, as there are not many
results for the exact bipartite entanglement of formation $E_{F}$ of
N-partite states. Lower bounds for $E_{F}$ of N-GHZ states are presented for
comparison with P1's results. We also show that proof of P1's optimality for
N-GHZ states would imply its optimality for Schmidt decomposable states as
well.

We consider some generalizations of protocol P1, which can be shown to be
more efficient than P1 for some classes of states (such as N-Toast states).
It would be interesting to investigate in more depth the bounds for $E_{F}$
obtained by the proposed generalizations.

We acknowledge support from the Royal Society and the Brazilian agency
Coordena\c{c}\~{a}o de Aperfei\c{c}oamento de Pessoal de Nivel Superior
(CAPES).

\bigskip

---------------------------------------------------------------------------------

[1] C. H. Bennett, G. Brassard, C. Cr\'{e}peau, R. Jozsa, A. Peres, and W.
K. Wootters, Phys. Rev. Lett. \textbf{70}, 1865 (1996).

[2] C. H. Bennett, S. J. Wiesner, Phys. Rev. Lett. \textbf{69}, 2881 (1992).

[3] C. H. Bennett and G. Brassard ``Quantum Cryptography: Public Key
Distribution and Coin Tossing'', Proceedings of IEEE International
Conference on Computer Systems and Signal Processing, Bangalore, India,
December 1984, pp 175-179; D. Deutsch, A. Ekert, R. Jozsa, C. Macchiavello,
S. Popescu, and A. Sanpera, Phys. Rev. Lett. \textbf{77}, 2818 (1996); 
\textit{erratum} \textbf{80}, 2022 (1998); D. Mayers, LANL e-print
quant-ph/9802025; H. -K. Lo, H. F. Chau, LANL e-print quant-ph/9803006.

[4] R. Cleve, H. Buhrman, LANL e-print quant-ph/9704026; H. Buhrman, R.
Cleve and W. van Dam, LANL e-print quant-ph/9705033.

[5] C. H. Bennett, H. J. Bernstein, S. Popescu, B. Shumacher, Phys. Rev. A 
\textbf{53}, 2046 (1996).

[6] H. -K. Lo, S. Popescu, LANL e-print quant-ph/9707038.

[7] C. H. Bennett, D. R\"{o}hrlich, S. Popescu, J. A. Smolin, A. V.
Thapliyal, in preparation.

[8] V. Vedral, M. B. Plenio, M. A. Rippin, P. L. Knight, Phys. Rev. Lett. 
\textbf{78}, 2275 (1997).

[9] B. Schumacher, Phys. Rev. A \textbf{51}, 2738 (1995).

[10] A. Zeilinger, M. A. Horne, H. Weinfurter, and M. \^{Z}ukowski, Phys.
Rev. Lett \textbf{78}, 3031 (1997).

[11] J. Kempe, LANL e-print quant-ph/9902036.

[12] B. Simon J. Math. Phys. \textbf{14}, 1940 (1973).  See also
G. Bacciagaluppi, M. J. Donald and P. E. Vermass, Helvetica Physica
Acta \textbf{68}, 679 (1995).

\end{document}